\documentclass[a4paper,prb,twocolumn,showpacs,preprintnumbers,amsmath,amssymb,groupeaddress,superscriptaddress]{revtex4}
\usepackage{graphicx}
\usepackage{dcolumn}
\usepackage{bm}
\usepackage{nicefrac} 
\usepackage{amsmath}
\usepackage{natbib}
\usepackage{scrextend} 

\newcommand{\abs}[1]{\left\vert {#1}\right\vert}

\def\d{\hbox{\rm d}}

\def\muB{\hbox{$\mu_B$}}

\newcommand{\commentout}[1]{}

\def\addrParma{Dipartimento di Scienze Matematiche, Fisiche ed Informatiche, Universit\`a  di Parma, Parco Area
  delle Scienze 7A, I-43100 Parma, Italy}
\def\addrDelft{Fundamental Aspects of Materials and Energy (FAME), Faculty of Applied Sciences, Delft University of Technology, Mekelweg 15, 2629 JB Delft, The Netherlands}
\def\addrPavia{Dipartimento di Fisica, Universit\`a di Pavia,
  Via Bassi 6, I-27100 Pavia, Italy}

\begin{document}

\title{Mixed magnetism, nanoscale electronic segregation and ubiquitous first order transitions in giant magnetocaloric 
  MnFeSiP alloys detected by $^{55}$Mn NMR}

\author{R. Hussain}
\altaffiliation{Present address: \addrPavia}  
\affiliation{\addrParma}
\author{F. Cugini}
\affiliation{\addrParma}
\author{S. Baldini}
\affiliation{\addrParma}
\author{G. Porcari}
\affiliation{\addrParma}
\author{N. Sarzi Amad\`e}
\affiliation{\addrParma}
\author{X. F. Miao}
\affiliation{\addrDelft}
\author{N. H. van Dijk}
\affiliation{\addrDelft}
\author{E. Br\"uck}
\affiliation{\addrDelft}
\author{M. Solzi}
\affiliation{\addrParma}
\author{R. De Renzi}
\affiliation{\addrParma}
\author{G. Allodi}
\email{Giuseppe.Allodi@fis.unipr.it}  
\affiliation{\addrParma}
\date{\today}

\begin{abstract}
  We report on a study on a representative set of Fe\textsubscript{2}P-based MnFePSi samples by means of \textsuperscript{55}Mn NMR  in both zero and applied magnetic field. The first-order nature of the magnetic transition is demonstrated by truncated order parameter curves with a large value of the local ordered moment at the Curie point, even at compositions  where the transition appears second order from magnetic measurements. No weak ferromagnetic order could be detected at Si-poor compositions showing the kinetic arrest phenomenon, but rather the phase separation of fully ferromagnetic domains from volume fractions where Mn spins are fluctuating. 
  The more pronounced decrease of the ordered moment at the $3f$ sites on approaching $T_C$, characteristic of the mixed magnetism of these materials, is demonstrated to be driven by a vanishing  
spin density rather than enhanced spin fluctuations at the $3f$-site.
An upper limit of $0.03$~\muB\  is set for the fluctuating Mn moment at the $3f$ site 
by the direct detection of a \textsuperscript{55}Mn NMR resonance peak  in the Mn-rich samples above $T_C$, showing nearly temperature-independent frequency shifts. 
A sharper secondary peak observed at the same compositions reveals however the disproportionation of a significant $3f$-Mn fraction with negligible hyperfine couplings, which retains its {\em diamagnetic } character across the transition, down to the lowest available temperatures. Such a diamagnetic fraction qualitatively accounts for the reduced average $3f$ moment previously reported at large Mn concentrations.
\end{abstract}

\pacs{76.60.-k, 
  75.30.Kz, 
  75.30.Sg, 
74.62.Dh  
}

\maketitle
\section{Introduction}

Magnetic materials showing a first order magnetic transition (FOMT) have been attracting sustained research interest because of their inherent giant magnetocaloric effect, originating from the large entropy change taking place 
at the transition.  The latter is the key to their employment for magnetic refrigeration, whereby the vapor based thermodynamic cycles of traditional refrigerators are replaced by magnetization-demagnetization cycles, which are environment-safer and  potentially more efficient. To this end,  Fe$_2$P-based
Mn$_{x}$Fe$_{1.95-x}$Si$_y$P$_{1-y}$ alloys ($1 \le x < 1.95$)
are among the most promising materials. 
Their paramagnetic (PM) to ferromagnetic (FM) FOMT is governed by 
a magneto-elastic transition taking place at the Curie point, characterized by a sizable variation of the crystal cell parameters without a symmetry change. By varying the composition,  $T_C$ can be tuned over a wide temperature interval including room temperature, which makes these systems eligible for real case applications.

Like their parent Fe$_2$P compound, the
 Mn-Fe-Si-P systems
 crystallize in the hexagonal space group P6$\bar{2}$m, with a crystal structure characterized by the stacking of $3g$, $1b$ sites and  $3f$, $2c$ sites on alternate layers. The magnetic Fe, Mn ions occupy the $3f$ and $3g$ sites, with a marked preference of Mn for $3g$ site \cite{dung2011, dung2012}, while  the $1b$ and $2c$ sites are occupied by the non metallic P and Si ions, with a partial preference of 
 Si for $2c$ sites at 
 $y > 1/3$.  \cite{miao2014}
The magnetism of the $3f$ ions exhibits an itinerant character witnessed by a fractional magnetic moment $\le 1.5$ \muB\ at low temperature, as compared to
$\ge 2.5$~\muB\ on Mn at the $3g$ sites, which behave as nearly localized spins.
The moment at the $3f$ sites exhibits a steep drop at $T_C$, in coincidence with a marked in-plane lattice contraction, partly compensated by an expansion along $c$. The accompanying magneto-elastic transition at $T_C$
 fits into the scenario of an enhanced chemical bonding of the $3f$ atoms, and increased spin itineracy above the FOMT.
The coexistence of large  $3g$ Mn moments with weaker and possibly vanishing $3f$ moments is referred to as mixed magnetism. However, the exact nature of the electronic state of the $3f$ ions in the PM state is still controversial.
X-ray magnetic circular dichroism (XMCD) experiments indicated in fact that the moment quenching above $T_C$ is incomplete, \cite{yibole2015}
in contrast with earlier predictions for  a total $3f$ moment extinction by band structure calculations. \cite{advmat2011}

The  critical temperature and the character of the magneto-elastic transition are strongly affected by the  Mn/Fe and Si/P substitutions. An increasing substitution of P with the larger Si ions at the $2c$ sites gives rise to an in-plane expansion, accompanied by increased $3f$-$3f$ and decreased $3f$-$3g$ minimum distances. In agreement with the tendency observed on crossing the FOMT, this leads to a relative localization of the $3f$ moments and an enhanced inter-layer exchange coupling, hence a more robust ferromagnetism and an increased $T_C$ are observed. Silicon-poor compositions, on the other hand, yield lower $T_C$, an unsaturated magnetic moment and strong magnetization hysteresis vs.\ both temperature and applied field. The latter are manifestations of a {\em kinetically arrested} magnetic transition, i.e.\ a frozen-in metastable state separated from thermodynamic equilibrium by a large free-energy barrier. The kinetic arrest can be however mitigated and eventually suppressed  by doping a few percent boron into the $1b$ sites, producing a qualitatively similar, though more dramatic effect as the Si/P substitution.
The progressive substitution of Fe by Mn at the $3f$ sites at Mn concentrations $x > 1$, on the other hand, depresses $T_C$ and drives a decrease of the ordered moment. \cite{dung2012}
However, the magnetoelastic transition coupled to the FOMT is depressed as well, as witnessed by a smaller step in the lattice parameters. As a consequence, the first order character of the magnetic transition becomes less pronounced.

\begin{table}
  \caption[]{Composition and critical temperatures of the investigated samples.   }
  \label{tab:samples}
\begin{ruledtabular}
\begin{tabular}{llll} 
  Sample       & Composition$_{\LARGE \phantom A}$     &  \mbox{$T_C^{\uparrow}$ (K)\footnotemark[1]}  &  \mbox{$T_C^{\downarrow}$ (K)\footnotemark[2]} \\
  
  \hline

S1$^{\large \phantom A}$  & Mn$_{1.27}$Fe$_{0.68}$Si$_{0.52}$P$_{0.48}$ &   286.6(2) &  285.2(2) \\

S2 & Mn$_{1.7}$Fe$_{0.25}$Si$_{0.5}$P$_{0.5}$ &   178.5(3) &  175.5(3) \\

S3 & MnFe$_{0.95}$Si$_{0.29}$P$_{0.71}$ &   199(1) &  73(3) \\
S4 & MnFe$_{0.95}$Si$_{0.33}$P$_{0.67}$B$_{0.03}$  &   261.7(3) &  242(1) \\
\end{tabular}
\end{ruledtabular}
\footnotetext[1]{From $M(T)$ on warming (Fig.~{\protect \ref{fig:MagneticMvT}}) .}
\footnotetext[2]{From $M(T)$ on cooling (Fig.~{\protect \ref{fig:MagneticMvT}}).
}
\end{table}

In this paper, we address the study of the magnetic transitions and of 
mixed magnetism  in a representative set of Mn-Fe-Si-P compounds by means of $^{55}$Mn nuclear magnetic resonance (NMR). In zero applied field (ZF), $^{55}$Mn nuclei resonate in a hyperfine field 
proportional to the ordered component of the on-site Mn electronic spin,
thus probing locally the magnetic order parameter and the electronic configuration of the Mn ions. 
$^{55}$Mn NMR provides complementary information to  $^{57}$Fe M\"ossbauer spectroscopy, 
\cite{mossbauer}
with a benefit for NMR due to its simpler spectra, yielding a precise determination of the order parameter even in the presence of a broad distribution of hyperfine fields.
As seen by a local probe of magnetism, a FOMT appears as a {\em truncated } transition, \cite{allodi_manganites}
i.e.\ a non-vanishing local moment probed by the hyperfine field  and, rather,
a phase-separated magnetic state with a vanishing volume of the ordered phase at the transition temperature.
The first-order nature of the magnetic transitions in Mn-Fe-Si-P compounds is established by $^{55}$Mn NMR even at Mn-rich compositions where a nearly second-order magnetic transition (SOMT) was inferred from magnetization and magneto-caloric effect measurements. The mixed magnetism of these systems is 
unambiguously ascribed 
to the vanishing of the spin polarization at the $3f$ sites in the PM phase, with a very stringent limit for the 
fluctuating $3f$ moment set by the comparison of NMR and magnetization data.
Moreover, a sizable fraction of diamagnetic manganese, independent of temperature, was found at the $3f$-site at $x>1$ compositions. Such a diamagnetic fraction, never reported earlier, seemingly plays a role in the decreasing average  $3f$ moment at increasing $x$ reported by neutron scattering and the lower $T_C$ of the  Mn-rich compounds.


\section{Experimental details}
\label{sec:experiment}

\subsection{Sample preparation and characterization}
\label{sec:experiment.samples}

\begin{figure}[t]
\centering
 \includegraphics[width=0.92\columnwidth]{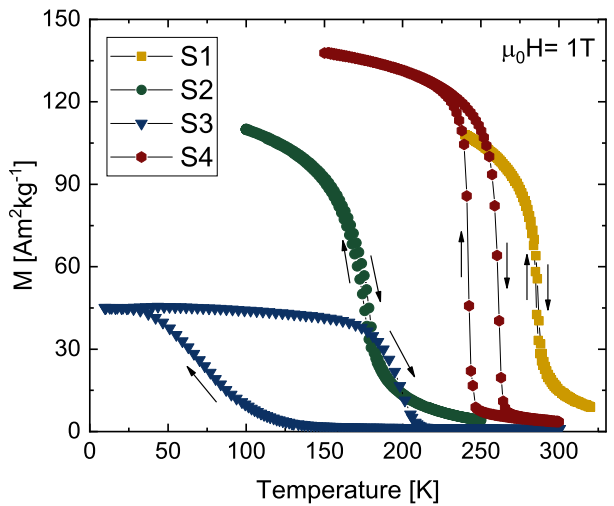}
 \caption{Magnetization curves vs.\ 
   temperature of the four S1-S4 samples, measured in an applied field of 1~T on both cooling and warming.}
\label{fig:MagneticMvT}
\end{figure}

The Mn-Fe-Si-P(-B) samples were prepared by ball milling for 10 hours of the binary parent Fe$_2$P compound and high-purity elementary Mn, Fe, red-P, Si (and B, when appropriate), weighed in proper amounts. The resulting fine powder was then pressed into small tablets, sintered  for two hours at 1100~$^\circ\mathrm{C}$  and annealed for 20 hours at 850~$^\circ\mathrm{C}$ in Ar atmosphere, 
and finally cooled down quickly to room temperature,
as detailed elsewhere \cite{dung2011, miao2016, thang2017}. 
The synthesis was targeted at four compositions  representative of different regimes in the  magnetic response: Mn$_{1.27}$Fe$_{0.68}$Si$_{0.52}$P$_{0.48}$, Mn$_{1.7}$Fe$_{0.25}$Si$_{0.5}$P$_{0.5}$,  Mn$_{1}$Fe$_{0.95}$Si$_{0.29}$P$_{0.71}$ and  Mn$_{1}$Fe$_{0.95}$Si$_{0.33}$P$_{0.67}$B$_{0.03}$, referred to hereafter as S1-S4, respectively.
Powders from the same batches of samples S2-S4 have also been the subject of the investigations published in Refs.~\onlinecite{miao2016,miao2016JSAMD,miaomiao2016}.
The sample homogeneity was assessed by powder X-ray diffraction, revealing a single crystal phase belonging to the appropriate  P6$\bar{2}$m space group.

The magnetic characterization of the samples was carried out by  a superconducting quantum interference device (SQUID) magnetometer (Quantum Design MPMS-XL) in the 2-350~K temperature range and by a MANICS DSM-8 magnetometer equipped with an oven in the 300-800~K range. 
The magnetization curves of the four samples, recorded  as a function of temperature in an applied field of 1~T, are shown in Fig.\ \ref{fig:MagneticMvT}. Sample S1 shows a sharp transition close to room temperature with negligible hysteresis, making this compound of interest for applications. The Mn-richer S2 shows a reduced $T_C$ and a smoother $M(T)$ curve suggesting a SOMT. The Si-poor S3 exhibits a strong temperature hysteresis and an unsaturated magnetic moment characteristic of a kinetically arrested magnetic state. The full magnetization is recovered and thermal hysteresis is nearly suppressed in S4, similar to S3 but for a slight B doping.
The transition temperatures of the various samples, determined as the inflection points of $M(T)$, are summarized in Table~\ref{tab:samples}.


\subsection{NMR experiments}
\label{sec:experiment.nmr}


The  NMR 
experiments were carried out by means of a 
home-built phase-coherent spectrometer \cite{hyrespect} and a helium-flow 
(in the 5-80~K range) or a nitrogen-flow cryostat (70-360~K) on finely powdered samples, in order to maximize the penetration of the radiofrequency (rf) magnetic field.
 Whenever an intense signal with a large rf enhancement was present, namely in ZF or in a moderate external field applied as a perturbation of the much larger spontaneous hyperfine field, measurements were performed by using  a small
 coil ($\le 50$~nH) wound around the sample and terminated onto a 50~$\Omega$ resistor as a probe. The sensitivity penalty of a non-resonant circuit was in fact compensated by the enhancement of the NMR arising from the hyperfine coupling of nuclear and electronic magnetization, characteristic of ferromagnets, \cite{riedi_eta, turov}
 while the untuned probehead allowed automated frequency scans.
A conventional LC resonator was however employed for the weaker $3f$ resonance lines very close to $T_C$, in order to improve sensitivity, as well as in all  the other cases.  
 
The NMR spectra were recorded  
by means of a standard 
$P-\tau-P$  spin echoes pulse sequence, with equal rf pulses $P$ of intensity and duration optimized for maximum signal, and delay $\tau$ limited by the dead time of the apparatus. Each spin echo, representing one frequency point in the spectrum, was analyzed by taking the maximum magnitude of the Fourier-transformed 
signal, as detailed elsewhere.  \cite{allodi_manganites}


\begin{table}
  \caption[]{Low-temperature mean $^{55}$Mn ZF-NMR frequencies and relative amplitudes of the main peaks in the spectra of the various samples.  }
  \label{tab:lowTnmr}
\begin{ruledtabular}
\begin{tabular}{lllll} 
  Sample      & $\bar\nu_{3g}$  & $\bar\nu_{3f}$  & $A_{3f}/A_{3g}$  &  \mbox{$A_{3f}/A_{3g}$}\footnotemark[1] \\ 
 ~ &(MHz) & (MHz)& (measured) & (expected)  \\
\hline
 S1 & 289.4(2) & 166.2(2) & 0.32(3) & 0.27 \\

S2 & 258.0(2) & 154.2(2) & 0.50(4) & 0.7 \\

S3  & 320.0(1) & 224(1) & 0.04 (1) & 0\\
S4 & 312.6(2) & 216.2(5) & 0.10 (2)  & 0\\
\end{tabular}
\end{ruledtabular}
\footnotetext[1]{Assumed equal to $x-1$ (see text).
}
\end{table}



\section{Experimental results}
\label{sec:results} 


The $^{55}$Mn spectra detected at low-temperature in zero and in a perturbing field, their evolution with temperature, the distinct behavior of the spontaneous field at the two Mn sites on approaching $T_C$, and the detection of 
non-magnetic Mn fractions from $^{55}$Mn NMR in applied fields, 
are presented for clarity in separate subsections.  Each one addresses specific issues: the assignment of the spontaneous $^{55}$Mn NMR peaks to corresponding crystal sites and the local moment probed therein; the order of the magnetic transition and the kinetic arrest phenomenon in silicon-poor compounds; the nature of mixed magnetism; and the electronic state at the $3f$ site above $T_C$, respectively.

\subsection{Mn moment and site occupancy}
\label{sec:results.localmoment}

\begin{figure}
\includegraphics[width=\columnwidth]{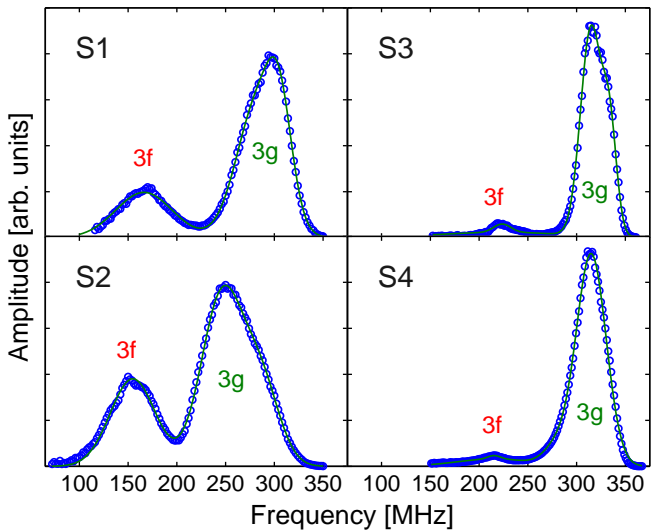}%
\caption{\label{fig:lowTspectra}
  ZF $^{55}$Mn NMR spectra of the S1-S4 samples at $T=5~K$. Spectral amplitudes are corrected for the frequency dependent sensitivity $\propto \nu^2$.
}
\end{figure}

Very intense spontaneous  $^{55}$Mn NMR signals were detected below $T_C$
by very low rf excitation power, thanks to
a sizable enhancement $\eta\approx  400$, \cite{riedi_eta}
in  the typical range of values for the coupling of nuclei to the electronic magnetization 
in domain walls.
The ZF spin echo amplitudes at low temperature  (5~K), divided by the
frequency-dependent sensitivity ($\propto \nu^2$),
are plotted  vs.\ frequency $\nu$  in Fig.\ \ref{fig:lowTspectra} for all our samples.
After such a correction, the plotted quantities correspond to the distributions of hyperfine fields at the  $^{55}$Mn nuclei. The spectra exhibit two broad peaks with composition-dependent positions and relative weights.   
The more intense resonance is found in the 260-320~MHz frequency interval (24-30~T in field units), whereas the minority peak, increasing in intensity with increasing Mn concentration, is located in the 150-220 MHz range (14-21~T). Assuming an
isotropic hyperfine coupling term in the order of -11~T/$\mu_B$ as in manganites \cite{allodi_manganites} and in other magnetic compounds, \cite{portis} and neglecting transferred contributions to the contact hyperfine field 
by neighboring ions, which are known to be very small in these systems, 
we estimate electronic spin moments of approximately 2.2-2.7~$\mu_B$ and  1.3-1.9~$\mu_B$ at the corresponding Mn sites. These values and the relative amplitudes of the two peaks are in qualitative agreement with the determination of the magnetic moment and the Mn occupancy at the two sites by neutron scattering. \cite{dung2012} The low- and high-frequency $^{55}$Mn resonance peaks are therefore unambiguously assigned to Mn nuclei at the $3f$ and $3g$ sites, respectively.

The best-fit mean frequencies of the two resonance peaks and the ratios of their integrated amplitudes, corresponding to the relative occupancy of the $3f$ and $3g$ sites by Mn, are listed for each sample in Table \ref{tab:lowTnmr}. 
The amplitude ratios are compared with the values expected for a total preference of Mn for the $3g$ site, whereby only Mn atoms in excess of 1 per formula unit (FU) occupy $3f$ sites. It is apparent from samples S3 and S4 ($x=1$), showing however a small but finite $3f$-Mn fraction, that the preferential occupation of $3g$ sites by Mn is not perfect, and some spillage of Mn and Fe to the disfavored sites occur. The Mn-rich sample S2 ($x=1.7$), on the other hand, exhibits a low-moment $3f$-Mn fraction significantly smaller than the expected 0.7 value.
The missing $3f$-Mn ordered fraction in this sample will be commented again in 
Sec.~\ref{sec:discussion},
along with the detection of a sizable amount of non-magnetic Mn ions.

\begin{figure}[t]
\includegraphics[width=\columnwidth]{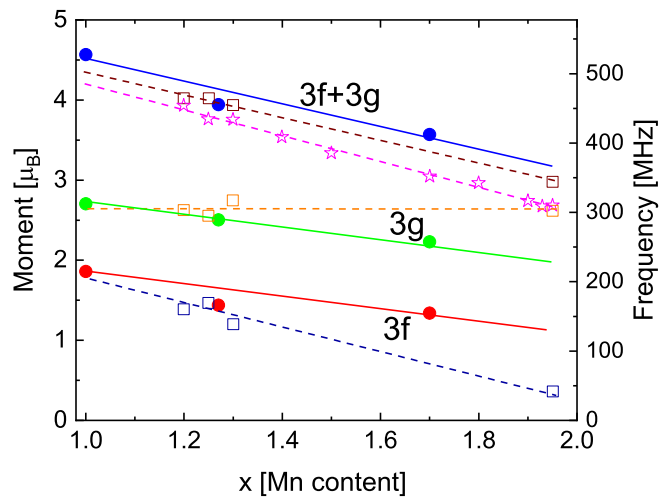}%
  \caption{\label{fig:NMR_neutrons}
Low-temperature  $3f$,  $3g$ and overall ordered moment per FU vs.\ Mn concentration $x$, determined by neutron scattering  (squares)  and macroscopic magnetization (stars) in various Mn-Fe-Si-P compounds from the literature, \protect{\cite{dung2012}} compared to the  $3f$ and  $3g$ $^{55}$Mn ZF NMR frequencies and their sum in the present samples (bullets).  NMR frequencies are scaled to moments by assuming a 11~T/$\mu_B$ hyperfine coupling constant.
}
\end{figure}

The mean spontaneous NMR frequencies of the $3f$ and $3g$ peaks are plotted vs.\ Mn concentration $x$ in Fig.~\ref{fig:NMR_neutrons}, overlaid to the $3f$, $3g$ and overall moment per FU determined by neutron scattering, and to the macroscopic saturation moment per FU reported in the literature for similar compounds. \cite{dung2012} In the figure, NMR frequencies are scaled to moments by assuming a mean hyperfine field of 11~T/$\mu_B$, as stated above. The total moment $\mu_{3g} + \mu_{3f}$ assessed by $^{55}$Mn ZF-NMR is in good agreement with its determinations by the other techniques. The agreement of the $\bar\nu_{3g}$ and  $\bar\nu_{3f}$  NMR  frequencies with the individual $3g$ and $3f$  moment values refined by neutrons is however poorer. In particular, NMR shows a similar relative decrease with increasing $x$ for the $\bar\nu_{3g}$ and  $\bar\nu_{3f}$ resonance frequencies, while a constant $\mu_{3g}$ and a more rapidly decreasing $ \mu_{3f}$ moment were estimated from neutron data.

\begin{figure}[t]
\includegraphics[width=\columnwidth]{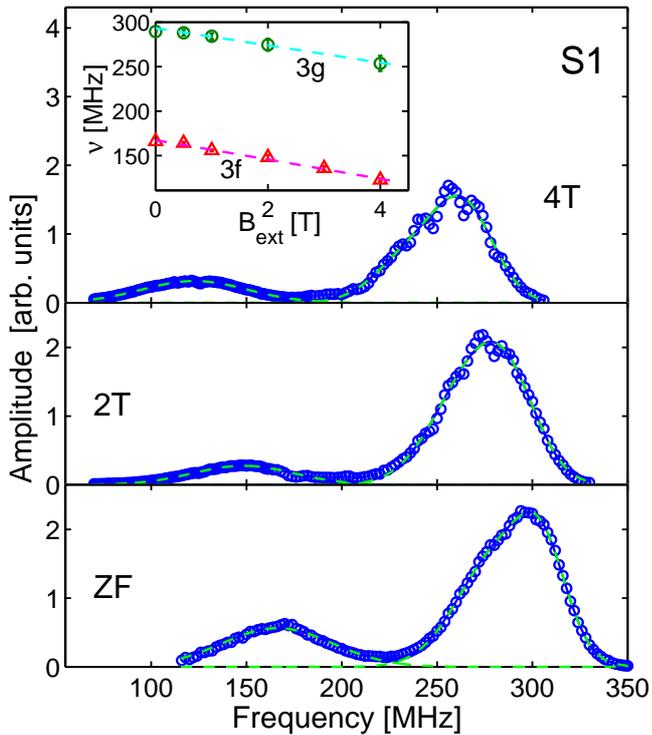}%
  \caption{\label{fig:NMRinfield_S1}
$^{55}$Mn NMR spectra of sample S1 at $T=5$~K, in zero and in perturbing applied fields. Inset: mean positions of the $3f$ and $3g$ lines as a function of the external field. The dashed lines are fits to Eq. \ref{eq:gammaline}.
}
\end{figure}

The attribution of both the resonance peaks to $^{55}$Mn nuclei (rather than $^{31}$P, which also experience a large transferred hyperfine field in the parent Fe$_2$P compound) \cite{f2p_NMR}
is further demonstrated by the application of moderate external fields acting as a perturbation of the much larger internal field. Typical spectra in external fields up to a few tesla are shown in Fig.~\ref{fig:NMRinfield_S1} for a representative sample (S1). Both the $3g$ and $3f$ mean resonance frequencies shift to lower frequency  following a linear dependence on the the applied field
\begin{equation}
    \bar\nu_{\alpha}(H) =\frac{\gamma}{2\pi} \abs{ B_{hf}^{(\alpha)}  + \mu_0 H }
  = \frac{\gamma}{2\pi} \left(\abs{ B_{hf}^{(\alpha)}}  - \mu_0 H \right )
  \label{eq:gammaline}
\end{equation}

($\alpha$ = $3g$, $3f$), where $B_{hf}^{(\alpha)}$ is the isotropic component of the
hyperfine field and $\gamma/2\pi = 10.5$~MHz/T is the gyromagnetic ratio of $^{55}$Mn (figure inset).
The collinear composition of the isotropic hyperfine field (collinear in turn to the on-site electronic moment) and  the external field is demonstrated
by the shift with an absolute rate $\mu_0^{-1} \abs{\d \bar\nu_\alpha / \d H}$
equal to the full $^{55}$Mn gyromagnetic ratio $\gamma/2\pi$,  without line splitting or appreciable broadening.
The negative sign of $d\nu_{\alpha} / d H$ indicates that the hyperfine field is  antiparallel to the electronic spin, in agreement with the core-polarization mechanism dominant in the contact hyperfine coupling of transition metal ions. \cite{freeman_watson}
All these facts prove therefore the full FM order of both sublattices.

\begin{figure}
\includegraphics[width=\columnwidth]{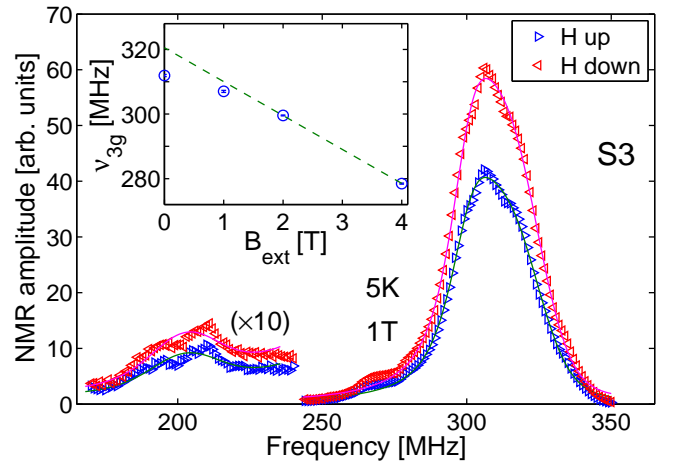}%
  \caption{\label{fig:NMRinfield_S3}
    $^{55}$Mn NMR spectra of sample S3 at $T=5$~K in an applied field 
    $B_{ext}=1$~T, applied after ZFC to the working temperature (blue right triangles) and after ZFC to 5~K, followed by an isothermal field cycle $0 \to 5~\mathrm{T}   \to 1~\mathrm{T}$  (red left triangles). Inset: mean position $\bar\nu_{3g}$ of the $3g$ line vs.\ $B_{ext}$. The dashed line a fit of the two highest-field points  $\bar\nu_{3g}(B_{ext})$ to Eq. \ref{eq:gammaline}, with $\gamma$ constrained to the $^{55}$Mn value.
}
\end{figure}

Similar field-dependent spectra were recorded in all samples.
In particular, they were detected also in the kinetically arrested S3, which shows a $M(H)$ curve saturating in two steps (Fig.~2b of Ref.~\onlinecite{miaomiao2016}), with the full saturation taking place in a field $H_s$ as high as $\mu_0H_s\approx5$~T, and a large hysteresis in $M(H)$ below $H_s$.\cite{miaomiao2016}
The field dependence of the mean frequency $\bar\nu_{3g}$ of the majority peak is plotted vs.\ field in the inset of Fig.~\ref{fig:NMRinfield_S3}, overlaid to a plot of the  $\nu(H)$ law  of Eq.~\ref{eq:gammaline}
(with slope $\gamma/2\pi$ set equal to the 10.5~MHz/T value of $^{55}$Mn).
It is apparent from the figure that, above an external field of 2~T corresponding to the saturation of the magnetically ordered domains, the collinear composition of the internal with the external field as of Eq.~\ref{eq:gammaline}
is obeyed by $\bar\nu_{3g}(H)$. The local magnetic order probed by NMR is therefore fully FM in sample S3 as well.
In the main panel of Fig.~\ref{fig:NMRinfield_S3} two low temperature spectra are compared, recorded in the same field $\mu_0H = 1~\mathrm{T} < \mu_0H_s$ applied after zero-field cooling (ZFC) and after an isothermal field ramp up to $H_s$ followed by a ramp down to the measuring field, respectively. Clearly, only the signal amplitude is hysteretic vs.\ field cycles, while the spectra are identical but for a vertical scaling relative to each other. This indicates that the microscopic properties of the magnetically ordered phase are unaffected by the applied fields, and only its volume is increased.
These results therefore provide evidence that the unsaturated magnetization of this sample is due to the coexistence of spatially segregated FM and non-FM domains, and that the full moment saturation develops as a field-induced nucleation of FM domains from the competing phase.

\subsection{First order transitions and kinetic arrest}
\label{sec:results.OMT}

\begin{figure}
\includegraphics[width=\columnwidth]{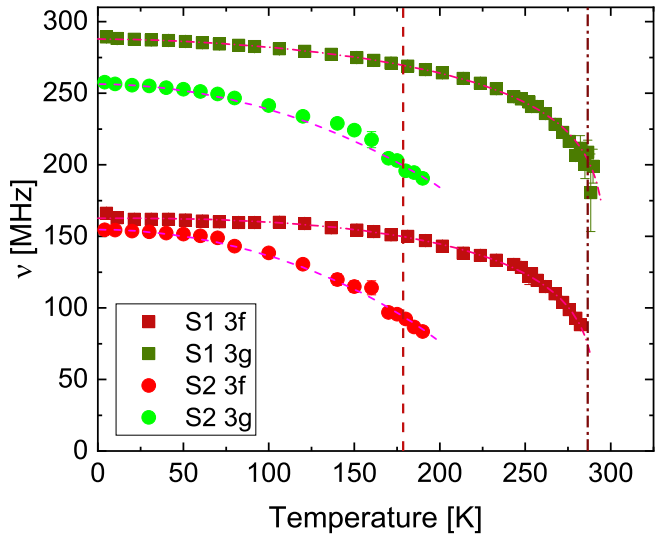}%
  \caption{\label{fig:S1S2_ordpar}
ZF $3g$ and $3f$ mean $^{55}$Mn resonance frequencies vs.\ temperature, recorded on warming in samples S1 (squares) and S2 (bullets). The lines overlaid to symbols are guides to the eye. Vertical dashed lines mark the upper transition temperatures $T_C^{\uparrow}$ of the two samples.
}
\end{figure}

\begin{figure}[t]
\includegraphics[width=\columnwidth]{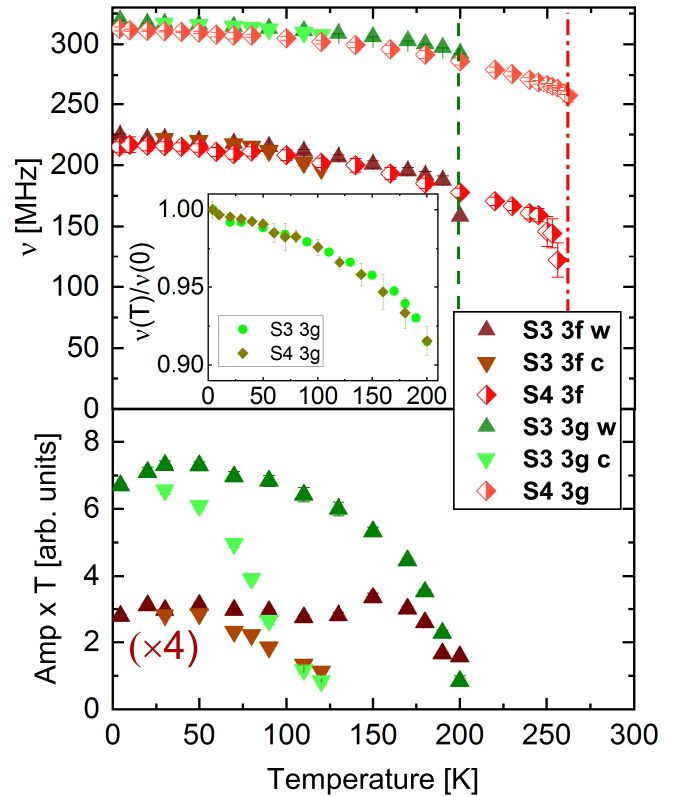}%
\caption{\label{fig:S3S4_ordpar}
  Top: ZF $3g$ and $3f$ mean $^{55}$Mn resonance frequencies vs.\ temperature of samples S3 on cooling (point-down triangles) 
  and warming (point-up triangles), 
  and of S4 on warming (diamonds). 
  The upper transition temperatures $T_C^{\uparrow}$ of the two samples are marked by vertical dashed line. Inset: normalized $3g$ resonance frequencies $\bar\nu_{3g}(T)/\bar\nu_{3g}(0)$ vs.\ $T$ of the two samples.
Bottom: integrated  $3f$ and $3g$ peak amplitudes times temperature $T$, as a function of $T$. For clarity, the $3f$ data are scaled by a vertical factor of 4.
}
\end{figure}

The 
ZF $^{55}$Mn NMR spectra could be recorded up to $T_C$ in  all samples. The temperature dependence of the centers of gravity $\bar\nu_{3g}$ and
$\bar\nu_{3f}$ of the two resonance peaks, proportional to the mean local moments at the two Mn sites, is plotted in Fig.~\ref{fig:S1S2_ordpar} and Fig.~\ref{fig:S3S4_ordpar} for samples S1-S2 and S3-S4, respectively. In the figures, the warming-up higher transition temperatures $T_C^\uparrow$  of Table \ref{tab:samples} are marked by vertical dashed lines for reference. It is apparent from the figures that
the  $\bar\nu_{3g}(T)$  order parameter curves do not vanish at $T_C$ in any sample. The reduced order parameter at the $3g$ site  $\bar\nu_{3g}(T)/\bar\nu_{3g}(0)$
is  approximately 0.8 at  $T_C$ and extrapolates to zero at a significantly higher temperature.
This holds true also for samples S2, showing a comparatively small thermal hysteresis and a magnetization vs.\ temperature dependence resembling that of a SOMT (Fig.~\ref{fig:MagneticMvT}). 
These {\em truncated} order parameter curves (i.e.\ not vanishing at $T_C$),  as seen by a local probe in direct space like NMR, indicate that the magnetically ordered phase does not collapse on warming due to critical spin fluctuations  but, rather, to an independent mechanism which abruptly breaks down exchange coupling.
The truncation effect is therefore a clear indication of a FOMT.
In Fe$_2$P-based alloys, the driving mechanism for the magnetic transition is known to reside in the concomitant magneto-elastic transition at $T_C$.
Similar truncated transitions have been observed e.g.\ in manganites, \cite{allodi_manganites, allodi_lpsmo} where they have been ascribed to the breakdown of half-metallicity and the onset of a polaron phase.  \cite{lynn}

In order to reconcile the seeming SOMT from magnetization data  in sample S2 with the marked first order character apparent from $^{55}$Mn NMR, it is worth noting that spontaneous $^{55}$Mn resonances are detected in this sample several kelvin above $T_C$  without noticeable anomalies in $\bar\nu_{3g}(T)$ and $\bar\nu_{3f}(T)$ (Fig.~\ref{fig:S1S2_ordpar}). This behavior indicates the coexistence of a FM and a PM phase over a wide temperature interval across the transition, with the FM fraction surviving above $T_C$ as a minority phase. Therefore, it is the volume fraction of the ordered phase which tends continuously to zero on warming, as a consequence of a broad distribution of critical temperatures for the driving magneto-elastic transition. This interpretation is actually corroborated by X-ray diffraction data, showing a smoother variation of the lattice parameters in this sample, as compared to the step-like jump observed in the other ones.\cite{miao2016}

The $3g$ and $3f$ order parameter curves of sample S3, showing a marked thermal hysteresis and the kinetic arrest phenomenon, are plotted vs.\ temperature on  both cooling and warming in Fig.~\ref{fig:S3S4_ordpar} (top panel). Very small and negligibly small differences in the values of  $\bar\nu_{3f}$  and  $\bar\nu_{3g}$, respectively, are detected between the cooling and warming scans, and only close to the lower transition $T_C^\downarrow$.
In the figure,  $\bar\nu_{3g}(T)$ and  $\bar\nu_{3f}(T)$ of S3 are overlaid  to the corresponding quantities of sample S4 containing the same nominal amount of Mn but showing in contrast higher $T_C$, a much narrower hysteresis and the full saturation moment. Notably, the spontaneous NMR frequencies of the two samples follow the same relative dependence of temperature and can be made to overlap  by a vertical scaling factor of $\approx 1.02$ (figure inset). The near coincidence 
of the $\bar\nu$ absolute values in these macroscopically very different samples is in essential agreement with the reported finding that the local $3g$ and $3f$ moments depend to leading order only on Mn content.\cite{dung2012,miao2014} Their slight difference is probably related to the larger amount of misplaced Mn atoms at the $3f$ sites found in S4 (Table \ref{tab:lowTnmr}).

The integrals of the NMR signal amplitude of S3 over the $3g$ and $3f$ peaks,  normalized by the sensitivity proportional to $\nu^2$ and the Boltzmann factor $1/T$, are plotted vs.\ $T$ in the bottom panel of Fig.~\ref{fig:S3S4_ordpar}.
Where the signal loss due to the spin-spin nuclear relaxation is negligible, i.e.\ far enough from
$T_C^\uparrow$, the plotted quantities are proportional to the number of resonating nuclei, hence to volume of the ordered phase.
In contrast to $\bar\nu_{3f}(T)$ and  $\bar\nu_{3g}(T)$, the normalized signal amplitude is strongly hysteretic vs.\ $T$, and its temperature behavior closely reproduces that of $M(T)$.
The thermal hysteresis of the magnetically ordered volume therefore agrees  with the onset of a net 
moment by the nucleation of fully FM domains, as also indicated by the  similar hysteresis vs.\ $H$ reported above.

Notably, a spin density wave (SDW) was detected in this sample by neutron scattering across $T_C$, and identified with the competing phase to the FM one. \cite{miaomiao2016}
No NMR signal component distinct from the two-peak spectrum of Fig.~\ref{fig:lowTspectra}c, however, could be detected down to the lowest temperatures. The missed detection of a SDW by our experiments may be due to a number of reasons, involving the different timescales of the two techniques as well as unfavorable experimental conditions for NMR. The SDW might be entirely dynamic in nature as sensed by the slower NMR probe (under this hypothesis, an upper limit $\tau_c \le 10^{-5}~\mathrm{s}$ may be set to its correlation time), making it indistinguishable from the PM phase. 
Even if, on the contrary, the SDW is mostly static, it might be subject to sizable and relatively slow ($\tau_c \ge 10^{-9}~\mathrm{s}$) spin or charge excitations, leading to exceedingly fast relaxations which wipe out
the NMR signal  of $^{55}$Mn nuclei therein. \cite{allodi_LSCO,allodi_wipeout}
 Finally, the detection of a SDW phase may be hindered by a diminished sensitivity of $^{55}$Mn NMR to non-FM signal components, in view of the smaller rf
 enhancement typically encountered in the latter.


\subsection{Mixed magnetism}
\label{sec:results.mixed}

\begin{figure}[t]
\includegraphics[width=\columnwidth]{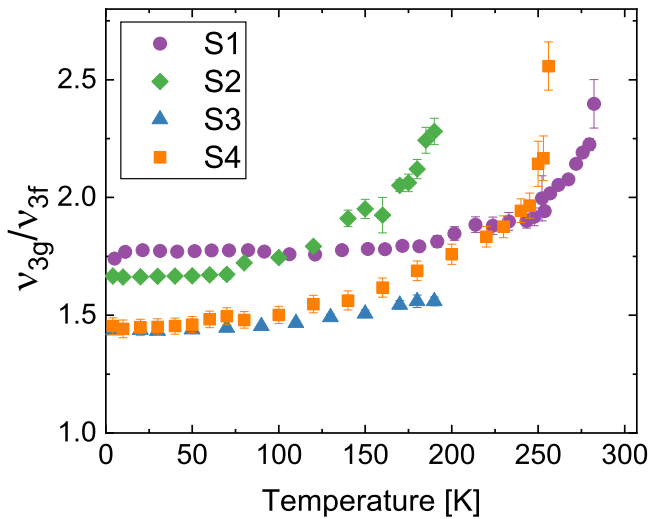}%
  \caption{\label{fig:3g_over_3f}
Mean frequency ratio $\bar\nu_{3g} /\bar\nu_{3f}$ of the $3g$ and $3f$ ZF $^{55}$Mn NMR peaks in the four samples, as a function of temperature.
}
\end{figure}

A careful examination of 
$\bar\nu_{3g}(T)$ and $\bar\nu_{3f}(T)$ in the various samples reveals that the two quantities, probing the ordered Mn moments at the $3g$ and the $3f$ sites,
do not scale with each other as a function of temperature and seemingly behave as independent order parameters.
In particular, $\bar\nu_{3f}(T)$ decreases faster than  $\bar\nu_{3g}(T)$ on warming. This is apparent from Fig.~\ref{fig:3g_over_3f}, showing a continuous increase of the $\bar\nu_{3g}(T)/ \bar\nu_{3f}(T)$  ratio  with temperature in all the four samples and a clear upturn at $T_C$ in samples S1, S2 and S4.
The steeper temperature dependence of the $3f$ ordered moment on approaching $T_C$, in agreement with previous reports by other techniques, \cite{dung2012, yibole2015}, \cite{advmat2011}  is a clear manifestation of the weaker magnetism of the $3f$ ion, possibly evolving into a non-magnetic $3f$ state. Its coexistence with the larger and more localized $3g$ moment is referred to altogether as {\it mixed magnetism}, and it is believed to play an important role in the FOMTs of this class of materials. \cite{miao2016ScrMat, miao2016JSAMD, miao2016, miaomiao2016}    
Nevertheless, it is 
clear from Figs.~\ref{fig:S1S2_ordpar} and \ref{fig:S3S4_ordpar} that also $\nu_{3f}(T)$, with the sole possible exception of sample S1, maintains a finite value at $T_C$, as in a truncated order parameter curve. This particularly evident in the kinetically arrested S3 sample, showing just a moderate increase with temperature in   $\bar\nu_{3g}(T)/ \bar\nu_{3f}(T)$ (Fig.~\ref{fig:3g_over_3f}). Such a quantitatively different behavior may be understood in view of the more marked truncation effect in this sample, whose mean hyperfine frequencies $\bar\nu_{3g}(T)$, $\bar\nu_{3f}(T)$ closely follow those of S4, in spite of the large difference in the transition temperatures of the two compounds. It seems therefore that the FOMT and the drop of the $3f$ moment are
independent phenomena,
especially
at compositions showing the kinetic arrest like that of S3, and that a non magnetic state of the $3f$ ion (discussed below) is achieved somewhere above
$T_C$.

\begin{figure}[t]
\includegraphics[width=\columnwidth]{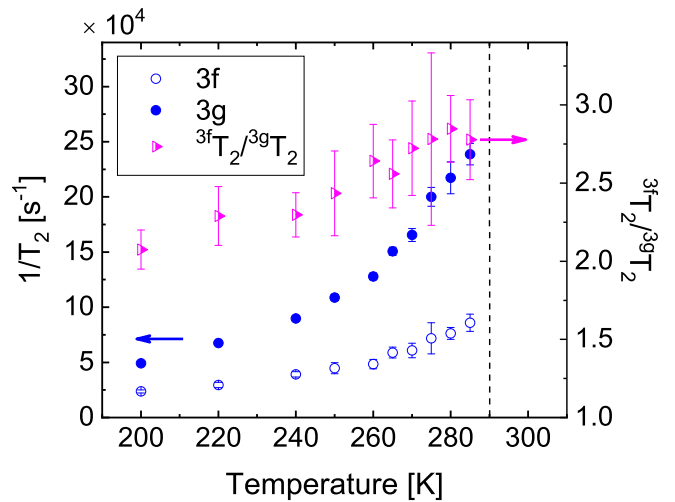}%
\caption{\label{fig:ZFrelax}
  Spin-spin $^{55}$Mn relaxation rates of sample S1, measured in ZF on the $3g$ (filled circles) and $3f$ peak (empty circles) as a function of temperature, and their ratios (triangles).
}
\end{figure}

The nature of the observed drop of the $3f$ {\em ordered} moment on approaching $T_C$ is demonstrated by nuclear relaxations. Spin-spin relaxation rates $T_2^{-1}$, measured in ZF on the two resonance peaks, are plotted vs.\ temperature in Fig.~\ref{fig:ZFrelax} for a representative sample (S1). 
Far enough from $T_C$, $T_2^{-1}$ is smaller at the $3f$ site by a factor of 2, in qualitative agreement with the smaller $\nu_{3f}$. This experimental ratio
is smaller than the $(\bar\nu_{3g}/\bar\nu_{3f})^2 \approx 3$ value expected from the isotropic hyperfine couplings of the two sites. Such a discrepancy is probably accounted for by anisotropic couplings, which also produce nuclear relaxations without a net effect on the mean resonance frequencies, and which are comparable for the two peaks as indicated by their similar absolute linewidths. Nevertheless, the  ratio of $T_2^{-1}$ at the $3g$ over the $3f$ site increases on warming. This rules out enhanced spin fluctuations (due e.g.\ to weakened exchange interactions) as the origin for the loss of the $3f$ ordered moment, as the latter would lead to relatively stronger nuclear relaxations on the $3f$ peak, contrary to evidence. The $3f$ moment drop is therefore due to the vanishing of the spin polarization and the tendency to a non-magnetic state of the $3f$ ions, in agreement with ab initio band calculations.   \cite{advmat2011}

\subsection{Vanishing $3f$ moment and diamagnetic Mn}
\label{sec:results.diaMn}

\begin{figure}[t]
\includegraphics[width=\columnwidth]{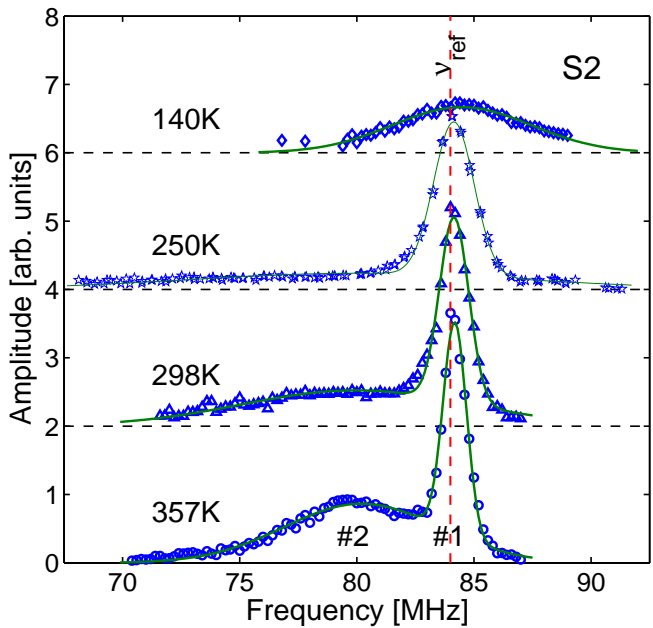}%
  \caption{\label{fig:PMspectra}
$^{55}$Mn NMR spectra of sample S2 in an applied field of 7.96~T ($^{55}$Mn reference frequency $\nu_{\mathrm{ref}}=83.998$~MHz) at several temperatures.  
}
\end{figure}

The PM phase at temperatures well above $T_C$ was investigated by $^{55}$Mn NMR in an applied field $B_{ext}=7.96$~T mostly in sample S2, featuring both a large $3f$ fraction and relatively low $T_C$ which matches the temperature range of our N$_2$-flow cryostat. A set of $^{55}$Mn NMR spectra recorded at several temperatures above $T_C$ is shown in Fig.~\ref{fig:PMspectra}. The highest temperature spectrum (357~K) consists of two well resolved resonance peaks: a sharper line with a small positive shift with respect to the $^{55}$Mn reference (83.998~MHz), referred to hereafter as line 1, and a much broader resonance with a negative shift $K\approx -4.5\%$ ($-4$~MHz in absolute units), referred to as peak 2.
The two peaks, whose intensity is not corrected for spin-spin relaxations in the plot, exhibit comparable raw integrated amplitudes.
However, line 2 is strongly relaxed, and its measured transverse relaxation time $T_2=12(1)~\mu\mathrm{s}$ is shorter than the dead-time-limited duration 
of the spin-echo pulse sequence employed to excite the resonance. The signal loss due to relaxation, on the contrary, is negligible for line 1 ($T_2=85(1)~\mu\mathrm{s}$). When the $T_2$ relaxation is accounted for, the corrected amplitude of peak 2 is estimated 4 times larger than that of line 1, which is therefore a minority signal.

\begin{figure}[t]
\includegraphics[width=\columnwidth]{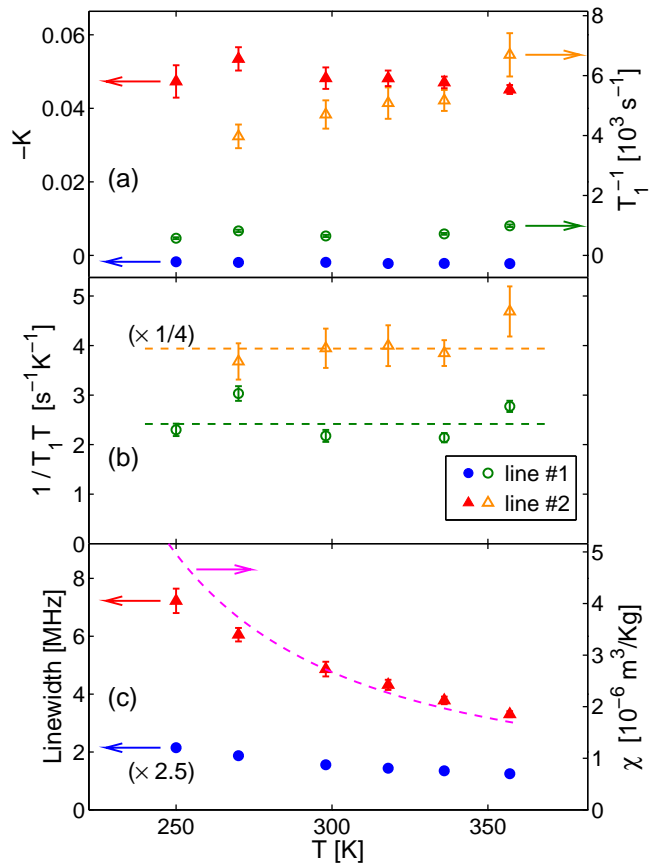}%
  \caption{\label{fig:PMdata}
Top: reverse-sign line shifts (filled symbols) and spin-lattice relations $T_1^{-1}$ (open symbols) as a function of temperature, for the two resonance lines in the $^{55}$Mn NMR  spectra of sample S2 in the PM phase (Fig.~\ref{fig:PMspectra}). Middle: Korringa product $T_1^{-1}T^{-1}$ vs.\ $T$ for the two  $^{55}$Mn resonance lines. For clarity, the plotted values of line 2 are divided by a constant ratio of 4.
Bottom: Gaussian NMR linewidths $\sigma$ vs.\ $T$ of the two  $^{55}$Mn resonance lines (symbols) and dc susceptibility measured in $\mu_0H=1$~T (line). For clarity, the $\sigma(T)$ data of the narrower line 1 are magnified by a constant factor 2.5 in the plot.}
\end{figure}

\begin{figure}[t]
\includegraphics[width=\columnwidth]{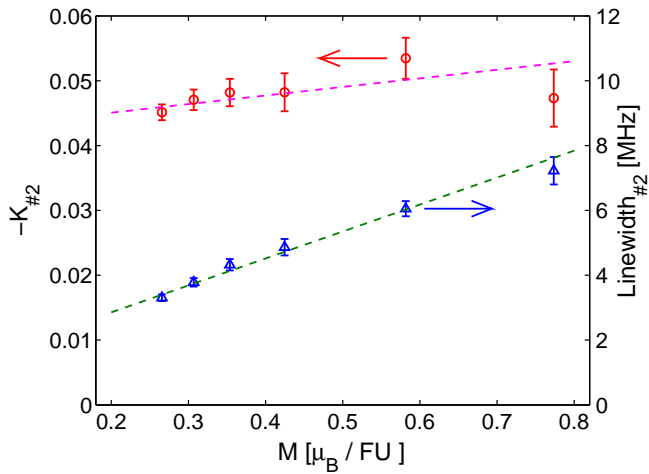}%
\caption{\label{fig:jaccarino}
  Negative lineshift $-K_{2}$ and linewidth $\sigma_{2}$ of line 2 in the $^{55}$Mn NMR  spectra of sample S2 of Fig.~\ref{fig:PMspectra}, plotted as a function of the macroscopic magnetic moment calculated from the $\chi(T)$ data of Fig.~\ref{fig:PMdata}.
}
\end{figure}

The evolution of the $^{55}$Mn NMR spectrum on cooling down to 250~K, a temperature still well above $T_C$, is summarized by Fig.~\ref{fig:PMdata}, showing the shifts $K(T)$ (panel a) and widths $\sigma(T)$ (c) of the two lines as a function of temperature. The spin lattice relaxation rates $T_1^{-1}(T)$ are also plotted in Fig.~\ref{fig:PMdata}a, overlaid to the $K(T)$ data.
On both peaks, they obey the Korringa law $T_1^{-1} \propto T$ typical of metals, \cite{abragam, slichter}
as it is apparent from Fig.~\ref{fig:PMdata}b, showing Korringa products $T_1^{-1} T^{-1}$ independent of $T$ within experimental accuracy.
The 
$T_1^{-1}$ rates of peak 2 are larger   
than on peak 1 by approximately a factor of 7, in agreement with
a similar ratio observed in $T_2^{-1}$  (see above). 
 This indicates, along with the larger linewidth of peak 2, that $^{55}$Mn nuclei therein experience stronger interactions with electronic spins.
The magnitude of its Gaussian linewidth $\sigma_{2}(T)$ 
is much larger than the dipolar contribution, for which a limiting value of approximately 2~MHz is calculated in the FM phase, while its temperature dependence mimics that of the of the magnetic susceptibility $\chi(T)$ (overlaid to the $\sigma(T)$ data of Fig.~\ref{fig:PMdata}c for reference). These two features of  $\sigma_{2}(T)$ demonstrate therefore a sizable {\em anisotropic} hyperfine coupling term for the corresponding nuclei. 
The isotropic hyperfine contact term proportional to the on-site spin density, on the other hand, is probed by the line shift $K(T)$ after the subtraction of possible temperature-independent chemical shift terms. Figure \ref{fig:PMdata}a clearly shows that the sizable shift $K_{2}(T)$ of line 2 is nearly independent of temperature, and it is therefore dominated by a chemical shift in the order of
-4\%. Such a large value demonstrates its origin from low-lying excited crystal field states via the van Vleck mechanism,\cite{vanvleck} and is  comparable to the  values found in the non-magnetic ground state of transition metal ions, like e.g.\ low-spin cobalt. \cite{allodi_LSCO}
The isotropic  hyperfine coupling of these $^{55}$Mn nuclei, probed by the residual temperature-dependent component of  $K_{2}(T)$,
can be better estimated by comparison with magnetometry data
in the $K$ vs.\ $M$ plot of Fig.~\ref{fig:jaccarino}, where temperature is the implicit parameter. The linear fit in the figure extrapolates to a maximal spin-dependent component in $\Delta K_{2}^{(sat)}=(\partial K_{2}/\partial M) M_s  = -0.04(2)$ for a saturation moment $ M_s\approx 3\muB$, i.e.\ a hyperfine field $B_{hf}^{(iso)} = \Delta K_{2}^{(sat)} B_{ext}=-300\pm 150~\mathrm{mT} $ in the virtual magnetically saturated state which would be attained either in much larger magnetic fields or at lower temperature while preserving however the high-temperature electronic and structural properties of the material. Assuming an on-site origin for the electronic polarization at the nucleus, the latter corresponds to a saturation Mn moment of just $\approx 0.03~\mu_B$, according to the core-polarization coupling constant of about $11~\mathrm{T}/\mu_B$ demonstrated above by comparing NMR  with neutron scattering data. 
For reference, the extrapolation from the  analogous linear dependence of  $\sigma_{2}$ on $M$ (shown in Fig.~\ref{fig:jaccarino} as well) yields an asymptotic linewidth  $\sigma_{2}^{(sat)} = 25(2)~$MHz, whence a rms anisotropic hyperfine field $B_{hf}^{(anis)} = \sqrt{3}\,\sigma_{2}^{(sat)}2\pi/\gamma =4.1(3)$~T. Such a large value of  $B_{hf}^{(anis)}$ cannot arise from the tiny on-site spin moment estimated above,\cite{hfcoupling} and must be therefore a transferred contribution from neighboring magnetic ions. Given the presence of a large, though anisotropic, transferred hyperfine field at the corresponding Mn site, a transferred origin cannot be ruled out for its much smaller  $B_{hf}^{(iso)}$ as well. Therefore, this majority $^{55}$Mn NMR signal is compatible with a strictly spinless Mn species. In view of the mixed magnetism detected in these materials by several techniques, including the ZF $^{55}$Mn NMR experiments  on this exact sample reported in this paper, this resonance is undoubtedly assigned to Mn ions at $3f$ sites, and it proves that the drop of the $3f$ spontaneous resonance frequency observed below $T_C$ evolves into a  truly non magnetic state in the PM phase.
To this end, it is worth noting that the limiting linewidth $\sigma_{2}^{(sat)}$ of peak 2,  defined above, coincides numerically with the experimental linewidth of the $3f$ peak in the low temperature ZF spectrum (Fig.~\ref{fig:lowTspectra}).
This suggests a common origin from transferred hyperfine interactions, mostly unaffected by  the low-temperature transition of the $3f$ ions to a magnetic state, for the inhomogeneous broadening of line 2 and the ZF-$3f$ peak.

We now focus on the 
 sharper line 1, arising from a minority Mn fraction, as commented above.
It is apparent from  Fig.~\ref{fig:PMspectra} and \ref{fig:PMdata}a that its shift $K_{1}$, much smaller  than $K_{2}$  and opposite in sign,  is practically independent of temperature. Moreover the magnitude of its  linewidth $\sigma_{1}(T)$, also much smaller than  $\sigma_{2}(T)$, agrees with a dominant dipolar origin from classical electronic moments. These two facts indicate, respectively, vanishing isotropic and a small anisotropic hyperfine couplings,  i.e.\  a {\it diamagnetic} behavior. Indeed, this Mn species does not develop a magnetic ground state on cooling, unlike the majority fraction probed by peak 2, as it is witnessed by the detection of the corresponding resonance safely below $T_C$, broadened ($\sigma_{1}\approx 3$~MHz at $T=140$~K), but essentially unshifted (Fig. \ref{fig:PMspectra}).
Though a minority diamagnetic Mn fraction, however, its belonging to a non magnetic impurity phase can be safely ruled out. Its amount is estimated in fact as large as $0.12(2)$ atoms per FU from the amplitude of its $^{55}$Mn NMR peak  relative to peak 2 probing $3f$ Mn. This figure also supported by its comparison with the intensity of $^{31}$P nuclear resonance in this sample, yielding the same estimate. \footnote{$^{31}$P NMR in these samples is the subject of a distinct investigation and will be reported elsewhere.} A spurious phase in such an amount would have been detected by X ray diffraction, contrary to experimental evidence. Moreover its nuclear relaxations 
are only moderately weaker than on peak 2 and 
in the typical range of nuclei of non magnetic ions in magnetic materials, like e.g.\ $^{139}$La in lanthanum manganites \cite{allodi_ca50, allodi_wipeout} and cobaltates.\cite{allodi_LSCO}
This proves that these nuclei experience electronic spin fluctuations and therefore belong to the proper MnFeSiP phase.

\begin{figure}[t]
\includegraphics[width=\columnwidth]{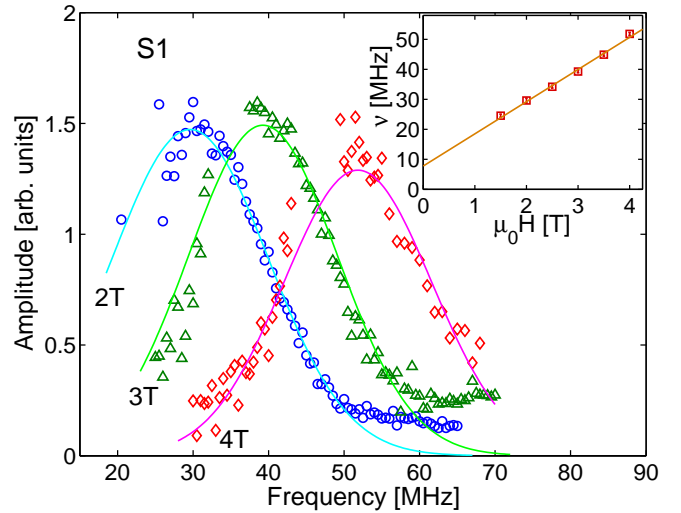}%
  \caption{\label{fig:S1_diaMn_5K}
$^{55}$Mn NMR resonance lines from the diamagnetic Mn fraction of sample S1 at $T=5$~K, in various applied fields. Inset: mean resonance frequency $\bar\nu$ vs.\ applied field.
}
\end{figure}

An indication on the origin of this so-called diamagnetic Mn fraction is provided by its systematic search in the other samples. 
Resonances perfectly similar to lines 1 and 2, except for a larger chemical shift of the former ($K_{1}\approx 1.3$) and overall broader spectra due the vicinity of $T_C$,  were found in sample S1 as well above 290~K up to he highest available temperature (360~K). In such a temperature range, the very large linewidth of peak 2 makes a reliable assessment of its position and amplitude, hence  a cross-calibration of the amplitudes of the two peaks, impossible. Qualitatively,
the intensity of peak 1 in S1  appears comparable or just slightly smaller than in S2. Its origin from diamagnetic Mn atoms, which retain their non-magnetic character down to the lowest temperature, is confirmed by
the $^{55}$Mn resonance peaks detected at $T=5$~K in the frequency scans of
Fig.~\ref{fig:S1_diaMn_5K} in moderate applied fields (so that the corresponding frequency intervals
are disjoint from
the field-shifted $3f$ spontaneous resonance). The common origin of these peaks with the high-temperature line 1
is proven by 
their shift with field 
according to the  $^{55}$Mn gyromagnetic ratio $\gamma$ (times a small correction for the chemical shift $K_{1}$) and a small positive internal field  $B_{hf}=0.7$~T, 
in agreement with the relation
\begin{equation}
    \bar\nu(H) =\frac{\gamma}{2\pi}  \left [ B_{hf}  + \mu_0 (1+K_{1})H \right ]
  \label{eq:gammaline_dia}
\end{equation}

(see Fig.~\ref{fig:S1_diaMn_5K} inset). The positive sign of $B_{hf}$, in contrast with
the negative spin hyperfine coupling of transition metal ions arising from core polarization, warrants in fact its transferred origin, 
compatibly with a spinless Mn state.

The search for signals from non-magnetic Mn at both room and low temperature, on the contrary, was unsuccessful in samples S3 and S4. Based on the estimated sensitivity of our NMR measurements, the missed detection of such signals poses an upper limit to a possible  diamagnetic Mn fraction not exceeding 0.01 Mn atoms per FU. 
Its practical absence in these two samples,  incidentally 
with the same $x=1$ Mn content and hence with a small, if not vanishing,
concentration of Mn atoms at the $3f$ sites (Tables \ref{tab:samples}, \ref{tab:lowTnmr}), as opposed to its sizable amount found in the Mn-richer S1 and S2 samples, suggests that diamagnetic Mn atoms are
located at $3f$ sites. 

\section{Discussion and conclusions}
\label{sec:discussion}

ZF $^{55}$Mn NMR proves that the  magnetic transition of Fe$_2$P based
Mn-Fe-Si-P alloys is strongly first order even at a Mn concentration as high as $x=1.7$ (sample S2), where a SOMT was inferred from a smoother $M(T)$ magnetization curve and a shallower magnetic entropy  peak at $T_C$.\cite{dung2011}
At this composition, however, the truncation effect on the magnetic order parameter, probed locally by the spontaneous $^{55}$Mn resonance frequency $\bar\nu_{3g}(T)$, is as severe as at other compositions. As already discussed above, the magnetic transition in this compound is actually characterized by a PM-FM phase coexistence over a wide temperature interval, indicative of a broad distribution of $T_C$ driven by a blurred magnetoelastic transition. 
Although the overall effect of such inhomogeneities on the macroscopic magnetism  accidentally mimics a SOMT, it is clear that here the continuous parameter at $T_C$ is the ordered volume instead of the local moment, and that this magnetic transition is not driven by critical fluctuations.
Based on this finding for this compound, whose macroscopic response shows the closest resemblance to that of a SOMT, we conclude that no real SOMT can be found in the composition phase diagram of Mn-Fe-Si-P alloys, and therefore FOMTs are a general property of these materials.

Similarly, a phase-separated state, rather than a weak FM structure,
is responsible for the  kinetically-arrested magnetic order of
the Si-poor S3 compound, apparent from its unsaturated and strongly hysteretic magnetization. $^{55}$Mn NMR demonstrates that the macroscopic magnetic moment develops by the nucleation of fully FM domains, while only the magnetically ordered volume is hysteretic vs.\ both temperature and applied field. The competing phase to the FM fraction, identified by neutron scattering with an incommensurate SDW, is possibly dynamic on the comparatively longer timescale of NMR so that it cannot be distinguished from the PM phase.

The mixed magnetism of the Mn-Fe-Si-P systems is detected by NMR through the independent temperature dependence of the  ordered moments at the $3f$ and $3g$ sites. The drop of the spin polarization at the $3f$ site on warming up to $T_C$, and not just its ordered component, is demonstrated by the steeper decrease of the $\bar\nu_{3g}$ NMR order parameter not accompanied by excess spin fluctuations, as it is witnessed by relatively weaker nuclear relaxations on the $3f$ resonance peak. The extinction of the $3f$ moment develops as a progressive reduction over a wide temperature range on warming and it is still incomplete at $T_C$, as it is apparent by the non-zero $\bar\nu_{3g}(T_C)$ value
(sample S1, featuring the highest $T_C$, is a possible exception to the
latter behavior).
This suggests that the FOMT in these materials, though related to the weakening of the $3f$ magnetism  observed with increasing temperature, is not simply triggered by it. 
This is particularly evident in the kinetically arrested S3 compound, showing only a moderate reduction of the $3f$ moment at its upper transition temperature  $T_C^\uparrow$. As already pointed out in Sec.~\ref{sec:results.mixed}, the stronger truncation effect on its order parameter 
curve, also involving a phase boundary with an incommensurate SDW phase,
might be the clue to the 
more pronounced
decoupling of the $3f$ moment quenching from the magneto-elastic transition in this sample.
Nevertheless, the drop of the $3f$ moment proceeds up to a spinless state of the $3f$ ions at temperatures high enough. 
This is verified directly  by $^{55}$Mn NMR  well above $T_C$ in a strong applied field whenever applicable, namely in the Mn-richer sample S2 (and, with much lower accuracy, S1) showing a significant fraction of $3f$ sites occupied by Mn ions. The direct detection of a non-magnetic $3f$ Mn state
confirms the initial prediction of a vanishing spin density at the $3f$ sites by band structure calculations, \cite{advmat2011} later challenged, however, by XMCD, \cite{yibole2015} thus solving a long-standing controversy.

Besides the majority fraction of ``normal'' $3f$ Mn atoms that develop itinerant magnetism below $T_C$, a Mn species with even weaker interactions with the surrounding magnetic ions and 
preserving its non-magnetic character down to the lowest temperatures, is revealed by the same NMR experiments  at the same Mn-rich compositions.
Though minority, such a 
diamagnetic fraction, 
reported in this work for the first time, is sizable ($\approx 0.12$ Mn atoms per FU in S2)  and $^{55}$Mn nuclei therein experience spin fluctuations.
It belongs therefore to the magnetic material under exam and not to an undetected (and unprecedented) non magnetic Mn impurity phase. 
Indeed, we are not aware of any non magnetic, ionic state of manganese, that are in contrast well documented e.g\. in cobalt, whose Co$^{3+}$ ion exhibits a low-spin (i.e.\ spinless) ground state in a octahedral crystal field. \cite{allodi_LSCO, allodi_caco} 
The disproportionation of Mn into high-spin and low-spin species at distinct sites, on the other hand, is possible in intermetallic compounds,
including metallic manganese itself ($\alpha$-Mn), \cite{alpha_Mn} whose Mn-IV atoms exhibit a nearly spinless state. \cite{alpha_Mn_NMR}
We argue therefore that the diamagnetic Mn state found in these compounds may be driven by a similar spontaneous electronic segregation within the proper Mn-Fe-Si-P phase. To this end,
it is worth noting that the Korringa behavior $T_1^{-1} \propto T$ obeyed by both $^{55}$Mn species (Fig.~\ref{fig:PMdata}b), indicating a dominant contribution to their spin-lattice relaxations by band electrons, agrees with the belonging to a metallic phase of the diamagnetic fraction as well. 
The lack of a significant diamagnetic Mn signal in the Mn-poorer S3 and S4 compounds where no Mn occupancy of $3f$ sites by Mn is expected (and a small one is detected by ZF NMR) indicates that diamagnetic Mn as well occupy $3f$ sites. In this view the electronic segregation, still to be identified, should involve two slightly different states of the $3f$ ion, weakly magnetic and strictly non magnetic, respectively.

The relevance of a diamagnetic Mn fraction at the $3f$ sites for the magnetic and magnetocaloric properties of the materials where it is present in a significant amount is unclear so far. However, it accounts qualitatively for some discrepancies found from the ZF $^{55}$Mn NMR spectra of the Mn-richer S2 compound ($x=1.7$), namely, a $3f$ to $3g$ peak intensity ratio smaller than expected (Table \ref{tab:lowTnmr}) and, on the contrary, a larger estimate for $3f$ moment than neutron scattering (Fig.~\ref{fig:NMR_neutrons}). It is noteworthy that the estimated diamagnetic fraction of sample S2 nearly equals within experimental errors its missing fraction of $3f$ ordered moments.
Sample S1 ($x=1.27$) also shows a significant diamagnetic Mn fraction. Here the $3f$ ZF NMR peak amplitude is in fair agreement with its expected value, but this may be the results of a compensation by off-site Mn  atoms spilled over from their preferential $3g$ site, as in sample S4. 
In the presence of a disproportionation between magnetic and non-magnetic species at the same crystallographic site, on the other hand, a probe in reciprocal space (neutron scattering) yields the average moment at the site,
while a probe in direct space (NMR)
selectively probes the moment of each magnetic species. This observation, therefore, qualitatively explains the steeper decrease of the $3f$ ordered moment  vs.\  Mn concentration  as determined by neutrons with respect to  NMR. We have however no explanation for the opposite relative dependence observed by the two techniques at the $3g$ site, where a moment decrease is detected by NMR only.


\section*{ACKNOWLEDGMENT}

This work is part of an Industrial Partnership
Program IPP I28 of the Stichting voor Fundamenteel
Onderzoek der Materie (FOM), which is financially supported
by the Nederlandse Organisatie voor Wetenschappelijk Onderzoek (NWO) and co-financed by BASF Future Business.



\bibliography{MnFeSiP}%
\end{document}